\pdfoutput=1
\documentclass{article}
\usepackage{ijcai19}

\usepackage{times}
\usepackage{soul}
\usepackage{url}
\usepackage[utf8]{inputenc}
\usepackage[small]{caption}
\usepackage{graphicx}
\usepackage{amsmath}
\usepackage{booktabs}
\usepackage{algorithm}
\usepackage{algorithmic}
\usepackage{xspace}
\usepackage{enumitem}
\usepackage{array}
\usepackage{amsfonts}
\usepackage{multirow}
\usepackage{subfigure}
\newcommand{\eat}[1]{}

\newcommand{\ie}{\emph{i.e.,}\xspace}

\newcommand{\baby}{\textsc{RNS}\xspace}

\title{A Review-Driven Neural Model for Sequential Recommendation}

\author{
Chenliang Li$^1$ \and
Xichuan Niu$^2$ \and
Xiangyang Luo$^3$\footnote{Xiangyang Luo is the corresponding author.} \and
Zhenzhong Chen$^{2}$ \And
Cong Quan$^4$
\affiliations
$^1$School of Cyber Science and Engineering, Wuhan University, China\\
$^2$School of Remote Sensing and Information Engineering, Wuhan University, China\\
$^3$State Key Lab of Mathematical Engineering and Advanced Computing, Zhengzhou, China\\
$^4$School of Computer Science, Wuhan University, China
\emails
\{cllee, niuxichuan\}@whu.edu.cn,
xiangyangluo@126.com,
\{zzchen, quancong\}@whu.edu.cn
}

\begin{document}

\maketitle

\begin{abstract}
Writing review for a purchased item is a unique channel to express a user's opinion in E-Commerce. Recently, many deep learning based solutions have been proposed by exploiting user reviews for rating prediction. In contrast, there has been few attempt to enlist the semantic signals covered by user reviews for the task of collaborative filtering. In this paper, we propose a novel \textbf{r}eview-driven \textbf{n}eural \textbf{s}equential recommendation model (named \baby) by considering users' intrinsic preference (long-term) and sequential patterns (short-term). In detail, \baby is devised to encode each user or item with the aspect-aware representations extracted from the reviews. Given a sequence of historical purchased items for a user, we devise a novel hierarchical attention over attention mechanism to capture sequential patterns at both union-level and individual-level. Extensive experiments on three real-world datasets of different domains demonstrate that \baby obtains significant performance improvement over uptodate state-of-the-art sequential recommendation models. 
\end{abstract}

%============================================
\section{Introduction}\label{sec:intro}
%============================================
Recommender systems are now an indispensable asset in many E-Commerce platforms to enhance their productivity. It is well recognized that these systems largely drive sales and improve user satisfaction by automatically pushing what a user is looking for~\cite{msq07:timothy}.
The next purchase decision made by a user is often influenced by her recent behaviors. For example, after buying a SLR camera, the user would be highly interested in camera lenses. Recently, sequential recommendation has drawn increasing attention from both academic and industrial circles. The task is to identify the next item that a user will purchase by considering her temporal preference as a sequence of purchased items.   

The key challenge of sequential recommendation is to dynamically approximate the current preference of a user by considering both the  general preferences and sequential patterns between items. As being the seminal work, FPMC extends the latent factor learning model by encoding the sequential transition patterns between items~\cite{www10:rendle}. The prediction made by FPMC is a linear combination of the user general preference and item sequential association which both are calculated by conducting inner-product between the corresponding latent vectors. Following this line, many variants are proposed to include more complex patterns~\cite{sigir15:wang,icdm16:he}. Recent advances in deep learning techniques have bred many neural solutions for sequential recommendation. Both convolutional neural networks (CNN) and recurrent neural networks (RNN) based models are proposed to capture a user's dynamic preferences~\cite{corr15:hidasi,sigir16:yu,wsdm18:chen,wsdm18:tang}, leading to state-of-the-art recommendation performance. However, these existing methods do not consider user reviews to enhance sequential recommendation. The semantic information contained in user reviews can reveal different features of items and also the preferences of users. It is widely recognized that exploiting review information would largely improve the rating prediction accuracy~\cite{kdd11:ctr,aaai14:topicmf,wsdm17:deepconn,cikm18:wu,tois19:wu,sigir19:li}.

To this end, we propose a \textbf{r}eview-driven \textbf{n}eural model for \textbf{s}equential recommendation, named \baby. The incorporation of review text brings more rich semantic information and strengthens model's expressive ability. Specifically, we first embed users and items into low-dimensional dense spaces through aspect-aware convolutional processing of the review documents, at the same time, user's general preference is extracted. Then a hierarchical attention over attention mechanism is employed to capture user sequential pattern at both union-level and individual-level. After that, we combine user's intrinsic (long-term) preference and temporal (short-term) preference to get a hybrid representation and make more accurate next items recommendation. On three real-world datasets from different domains, \baby outperforms the existing state-of-the-art methods remarkably. Overall, the key contributions of this work are summarized as below:

\begin{itemize}
\item[$\bullet$] We propose a novel review-driven neural model that exploits reviews for sequential recommendation. To the best of our knowledge, this is the first attempt to harness the rich semantics from reviews for this task.

\item[$\bullet$] We introduce an aspect-based convolutional network to identify user general preference from review document and utilize a hierarchical attention over attention mechanism to model user sequential preference at different granularities.

\item[$\bullet$] On three real-world datasets with diverse characteristics, our results demonstrate the superiority of the proposed \baby in terms of four metrics.
\end{itemize}
%============================================
\section{Related Work}\label{sec:related}
%============================================
In the following, we review the existing literatures that are highly relevant to our work: namely review-based recommendation and sequential recommendation.

%============================================
\subsection{Review-based Recommendation}
%============================================
Conventionally, many recommender systems are mainly developed in the paradigm of collaborative filtering (CF). Within this line of literatures, great effort is made under the framework of matrix factorization or the probabilistic counterpart~\cite{nips07:pmf,computer09:mf}. The idea is to represent each user or item as a low dimension latent vector. The binary interaction for a user-item pair is then estimated by performing inner-product of the corresponding latent vectors. However, the user-item interaction data is very sparse, which is an inherent obstacle for latent factor learning based solutions. Currently, to tackle the data sparsity problem, many techniques have been developed by utilizing the semantic signals provided by reviews. These advances have delivered state-of-the-art recommendation performance~\cite{kdd11:ctr,aaai14:topicmf,wsdm17:deepconn,recsys17:transnet,cikm18:wu,tois19:wu,sigir19:li}.

Many works propose to distill semantic latent factors from reviews with topic modeling techniques~\cite{jmlr03:lda}. These solutions incorporate the topical factors learnt from reviews into latent factor learning framework~\cite{kdd11:ctr,aaai14:topicmf}. However, given that the topic modeling represents a document as a bag-of-words (BOW), much semantic context information encoded with the word order is inevitably lost. The recent surge of deep learning has drawn remarkable attention from the community. Both convolutional neural networks (CNN)~\cite{emnlp14:kim} and recurrent neural network (RNN)~\cite{interspeech10:mikolov} are widely adopted to extract semantic representation from reviews for rating prediction~\cite{wsdm17:deepconn,recsys17:transnet,kdd18:mpcn,cikm18:wu,tois19:wu,sigir19:li}. Empowered by dense representation and neural composition ability, local contextual information in reviews is well preserved and easily composited to form high-level features. These neural solutions have delivered significant performance gain over the existing BOW-based alternatives. 

Many works also propose to derive aspect-level features from reviews.  Several approaches utilize the external toolkit for aspect extraction~\cite{sigir14:efm,cikm15:trirank}. Recently, several deep learning based solutions have been proposed. These works utilize attention mechanism to extract the aspect and derive the aspect-level representations~\cite{www18:alfm,cikm18:anr}. By highlighting the important words or aspect, the above mentioned solutions enable better rating prediction and facilitate semantic explainability. However, the aforementioned  techniques are mainly devised for the task of rating prediction. Neither of these works focuses on enhancing sequential recommendation with user reviews.

%============================================
\subsection{Sequential Recommendation}
%============================================
Sequential recommendation can be considered as a special kind of CF with implicit feedback. The effective learning of sequential patterns is widely verified to be a critical issue for sequential recommendation~\cite{is09:liu,www10:rendle,sigir15:wang,corr15:hidasi,icdm16:he,sigir16:yu,wsdm18:tang,wsdm18:chen,icdm18:kang}.
Earlier methods aim to extract sequential rules based on statistic co-occurrence pattern~\cite{icde95:agrawal,is09:liu}. FPMC~\cite{www10:rendle} is the first work that automatically encodes the sequential relations for items with latent vectors. Afterwards, several variants are proposed to incorporate auxiliary information. Fossil integrates high-order Markov chains and item similarity together for next item recommendation~\cite{icdm16:he}. HRM proposes a hierarchical architecture to capture both sequential patterns and users' general preference~\cite{sigir15:wang}. Recent efforts focus on enhancing sequential recommendation by using deep learning techniques, leading to the state-of-the-art performance. GRU4Rec and DREAM utilize a RNN network to capture sequential patterns~\cite{corr15:hidasi,sigir16:yu}. RUM utilizes the attention and memory mechanisms to express and update the users' preference in a dynamic fashion~\cite{wsdm18:chen}. The dynamic preference is modeled in both item and feature levels. Caser proposes to utilize both horizontal and vertical convolutional operations to harness the union-level and point-level sequential patterns~\cite{wsdm18:tang}. They also propose a skip behavior to consider the items in the next few actions for training. SASRec builds a self-attention model for sequential recommendation~\cite{icdm18:kang}. It can adaptively put different weights to historical items at each time step. Though these deep learning based solutions significantly enhance the performance of sequential recommendation. However, semantic information covered by user reviews are mainly overlooked. The proposed \baby in this work is the first attempt to exploit textual reviews in this field.

%============================================
\section{The Proposed Algorithm}\label{sec:algo}
%============================================
In this section, we firstly give a formal formulation of review-driven sequential recommendation problem and then present the details of our proposed \baby.

\begin{figure*}
\includegraphics[height=75mm, width=0.94\textwidth]{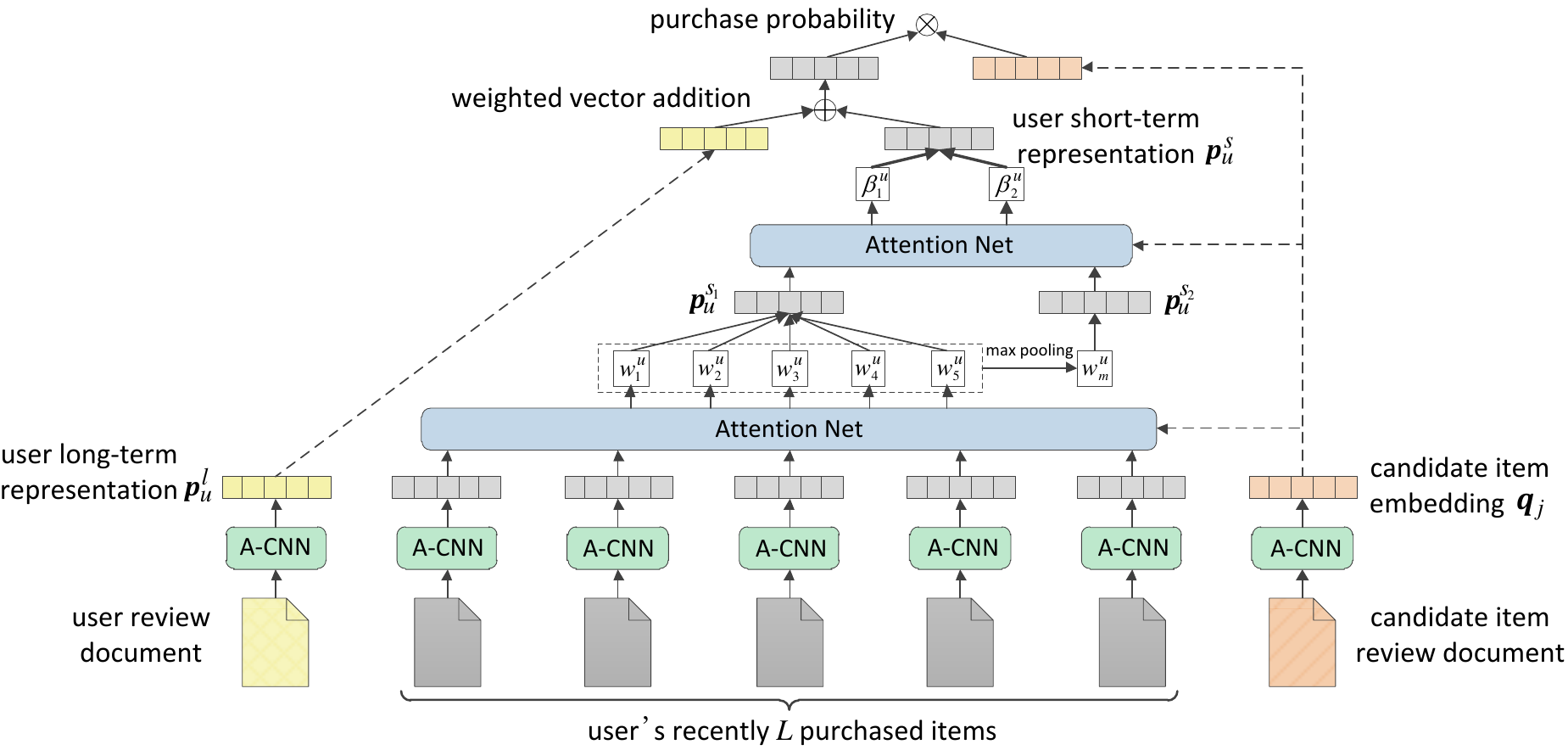}
\caption{The network architecture of \baby.}
\label{figure1}
\end{figure*}

%============================================
\subsection{Problem Formulation}
%============================================
Let $\mathcal{U}$ and $\mathcal{I}$ represent the user and item set, respectively. Each user $u \in \mathcal{U}$ is associated with a sequence of interacted items arranged chronologically as: $\mathcal{S}^u = (\mathcal{S}^u_1, \mathcal{S}^u_2, ..., \mathcal{S}^u_{|\mathcal{S}^u|})$, where $\mathcal{S}^u_t \in \mathcal{I}$ denotes the item purchased by user $u$ at time step $t$. In this work, we exploit user reviews to enhance sequential recommendation with implicit feedback. Consequently, we merge the set of reviews written by user $u$ and the set of reviews written for item $i$ to form user document $D_u$ and item document $D_i$ respectively. Given user $u$, her recently purchased $L$ items and their corresponding review documents, the goal is to rank candidate items in terms of the likelihood that the user will purchase in the next.

%============================================
\subsection{The Architecture of \baby}
%============================================
Figure~\ref{figure1} demonstrates the network architecture of \baby.
The objective of \baby is to derive the item representation, the user's intrinsic preference and temporal preference at union-level and individual-level from the corresponding reviews. Specifically, for each user $u$, we learn a long-term user representation $\textbf{p}^l_u$ by utilizing an aspect-aware convolutional network (A-CNN) over user document $D_u$. Likewise, each item $i$ is encoded as a representation vector $\textbf{q}_i$ via another parallel A-CNN over item document $D_i$. 

To capture user's temporal preference denoted by $\textbf{p}^s_u$, we utilize a hierarchical attention over attention mechanism to exploit the recent $L$ items (\ie $\textbf{q}^u_1,..., \textbf{q}^u_{L}$) she has purchased at both union-level and individual-level. Then, the user's current preference is formed through a linear fusion:
\begin{align}
\label{equ1}
\textbf{p}_u = \textbf{p}^l_u+\alpha\cdot\textbf{p}^s_u
\end{align}
where $\alpha$ controls the importance of the temporal preference. The preference score (\ie purchasing likelihood) of a candidate item $j$ is computed:
\begin{align}
\label{equ2}
s_{uj} = \sigma(\textbf{p}_u^\top\textbf{q}_j)
\end{align}
Here, we choose \textit{sigmoid} function $\sigma(x) = 1/(1+e^{-x})$ to meet the binary constraint (\ie 0/1). 

%============================================
\subsection{Aspect-Aware CNN (A-CNN)}
%============================================
Since the A-CNN architectures for user and item documents are identical (but with different parameters), we only describe the extraction process of user documents for simplicity. We first map each word in each user document into the corresponding embedding of $d$ dimensions. Thus, the user document is transformed to an embedding matrix $\textbf{M}_u\in\mathbb{R}^{l\times d}$, where $l$ is the document length. However, even for the same word, the semantics or sentimental polarity could be totally different for two different aspects in the same domain. For instance, word ``low'' in two sentences ``The price is very low'' and ``This computer is low resolution'' convey contrary sentiments towards aspects \textit{price} and \textit{resolution} respectively. So all words share the same $d$ dimensional vector across all aspects is unreasonable. Hence, we introduce aspect-specific embedding transformation matrix $\textbf{T}_a\in\mathbb{R}^{d\times d}$ and multiply it with original word embedding matrices:
$\textbf{M}^a_u=\textbf{M}_u\textbf{T}_a$
where $\textbf{M}^a_u$ is the embedding matrix for $D_u$ for aspect $a$. Therefore, we can represent $D_u$ as a tensor $\textbf{M}^*_u \in \mathbb{R}^{l\times d\times K}$ for $K$ aspects. We then utilize a multi-channel convolutional operation analogous to colored images for feature extraction. Specifically, A-CNN has $n$ filters $\textbf{F}^k\in\mathbb{R}^{h\times d\times K}, 1\leq k\leq n$, where $h$ is the filter height size that can have multiple values. For example, if $n=10,h=\{1,3,5,7,9\}$, then there will be two filters for each size. The $k^{th}$ filter derives its feature as follows:
\begin{align}
\textbf{z}^k[i] = ReLU(\textbf{M}^*_u[i:i+h-1] \odot \textbf{F}^k + b^k)
\end{align}
where $1\leq i\leq l-h+1$ is the start point of sliding window in the user document, $b^k$ is the bias, $\odot$ is the convolution operator and \textit{ReLU} is chosen as the activation function. The user's long-term preference is encoded as follows:
\begin{align}
\textbf{p}^l_u = [max(\textbf{z}^1), max(\textbf{z}^2),...,max(\textbf{z}^n)]
\end{align}
The item representation $\textbf{q}_i$ is extracted with the same procedure. 

Intuitively, the order of purchased items in user behaviors is pivotal for the sequential recommendation. Given a user $u$ and her recent $L$ purchased items $\textbf{q}^u_1,\textbf{q}^u_2,..., \textbf{q}^u_{L}$ arranged in chronological order, the latest item $\textbf{q}^u_{L}$ is more likely to reflect her temporal preference, while $\textbf{q}^u_1$ has minimal impact. Accordingly, we encode this temporal order information by including position embeddings to the item representation: $\textbf{q}^u_m=\textbf{q}^u_m+\textbf{o}_m$
where $\textbf{o}_m$ is embedding for $m$-th position, and $1\leq m\leq L$. Note that the sinusoid version of position embedding~\cite{nips:ashish} is also investigated, but it leads to unfavorable performance, which is in line with the observation in~\cite{icdm18:kang}.

%============================================
\subsection{Hierarchical Attention over Attention.}
%============================================
We further focus on deriving the user short-term preference $\textbf{p}^s_u$ from her recent $L$ items. A user's sequential preference could dynamically evolve over time. Moreover, different historical items could have different impacts on the user's next purchase decision. To accommodate these characteristics, we resort to the attention mechanism that has achieved great success in relevant works~\cite{kdd18:mpcn,tois19:wu}. Specifically, for each of user $u$'s $L$ purchased items $\textbf{q}^u_m (1\leq m\leq L)$ and each candidate item $\textbf{q}_j$, we first calculate the weight of each $\textbf{q}^u_m$ via an attention mechanism:
\begin{align}
w^u_m = softmax(\textbf{q}^\top_j \textbf{q}^u_m )=\frac{exp(\textbf{q}^\top_j \textbf{q}^u_m)}{\sum_{i=1}^{L}exp(\textbf{q}_j^\top \textbf{q}^u_i)}
\end{align}
Note that $\textbf{q}^u_m$ is a fusion of item representation and position embedding. This design enables \baby to easily incorporate temporal order for attention calculation. Then, the short-term preference of user $u$ with respect to candidate item $\textbf{q}_j$ is computed as a weighted sum of the item embeddings as follows:
\begin{align}
\label{equ9}
\textbf{p}^{s_1}_u = \sum_{m=1}^{L} w^u_m\cdot\textbf{q}^u_m
\end{align} 
Here, $\textbf{p}^{s_1}_u$ is user sequential preference representation at union-level. That is, $\textbf{p}^{s_1}_u$ is jointly encoded by all previous $L$ purchased items.  

We can simply take $\textbf{p}^{s_1}_u$ as the final short-term preference derived for user $u$, but it only captures sequential pattern at union-level, \ie every item of past $L$ ones has contribution for next item prediction. However, a user would buy an Apple earphone just because he bought an iPhone recently. In this case, other purchases except iPhone are just noises for making recommendation.
Hence, we further explore the previous purchase records' influence on future action at individual level, \ie identify the most relevant item. Inspired by the pointer mechanism~\cite{kdd18:mpcn}, we choose the item with maximum attention weight: 
\begin{align}
m_{u}&=\arg\max_m(w^u_m)\\
\mathbf{p}^{s_2}_u&=\mathbf{q}^u_{m_{u}} 
\end{align}
where $\textbf{p}^{s_2}_u$ is user sequential pattern representation at individual level. Then, we further utilize attention mechanism to discriminate which short-term preference is more important: union-level or individual-level. The final short-term user preference is calculated as follows: 
\begin{align}
\beta^u_n &= softmax(\textbf{q}^\top_j\textbf{p}^{s_n}_u)=\frac{exp(\textbf{q}^\top_j\textbf{p}^{s_n}_u)}{\sum_{i=1}^{2}exp(\textbf{q}_j^\top\textbf{p}^{s_i}_u)}\\	
\textbf{p}^s_u &= \sum_{n=1}^{2} \beta^u_n\cdot \textbf{p}^{s_n}_u
\end{align}
The preference score is then calculated according to Equation~\ref{equ1}-\ref{equ2}.

%============================================
\subsection{Model Inference and Optimization}
%============================================
The model parameters of \baby include word embeddings, position embeddings, aspect transformation matrices, and two A-CNN networks. 
For each user $u$, we extract each $L$ successive items and their next item as the target from the user’s sequence $\mathcal{S}^u$ to form a training instance. Following previous work~\cite{wsdm18:tang}, for each training instance $t_j$ with target item $j$, we randomly sample $x$ negative items, denoted as $\mathcal{N}(j)$. Let $\mathcal{C}^u$ be the collection of user $u$'s all training instances. The objective function is defined as binary cross-entropy loss with $L2$ norm regularization:
\begin{align}
\label{equ13}
\mathcal{L} = \sum_{u} \sum_{t_j \in \mathcal{C}^u} (-log&(s_{uj}) + \sum_{i \in \mathcal{N}(j)}-log(1-s_{ui})) + \lambda ||\boldsymbol{\Theta}||^2
\end{align}
where $\lambda$ is coefficient for the regularization and $\boldsymbol{\Theta}$ denotes all model parameters. The model is trained via Adma optimizer.
%============================================
\section{Experiments}\label{sec:exp}
%============================================
In this section, we conduct experiments on three real-world datasets to evaluate our proposed \baby against uptodate state-of-the-art methods. We then perform the ablation study to investigate the validity of each design choice made in \baby. Finally, we demonstrate the superiority of the proposed hierarchical attention over attention mechanism with case study.

%============================================
\subsection{Experimental Setup}
%============================================
\paragraph{Datasets.}

\begin{table}
	\centering
	\begin{tabular}{cccc}
		\toprule
		&IV&PS&THI\\
		\midrule
		\# users & 1,372 & 7,417 & 10,076 \\
		\# items & 7,957 & 33,798 & 66,710 \\
		\# interactions & 23,181 & 117,385 & 169,245\\
		sparsity & 99.79\% & 99.95\% & 99.97\%\\
		avg. actions per user & 16.9 & 15.8 & 16.8\\
		\bottomrule
	\end{tabular}
	\caption{Statistics of datasets}
	\label{tab1}
\end{table}

\begin{table*}
	\resizebox{\textwidth}{20.5mm}{
	\begin{tabular}{c|cccc|cccc|cccc}
		\toprule
		Datasets & \multicolumn{4}{c}{Instant Video} & \multicolumn{4}{c}{Pet Supplies} & \multicolumn{4}{c}{Tools and Home Improvement} \\
		\hline
		\midrule
		Measures@5 & Precision & Recall & NDCG & HR & Precision & Recall & NDCG & HR & Precision & Recall & NDCG & HR \\
		\hline
		Pop    		   & 0.0783&	0.0876&	0.0873&	0.2648&	0.0615&	0.0622&	0.0750&	0.2376&	0.0542&	0.0575&	0.0587&	0.2105 \\
		FPMC           & 0.1067&	0.1214&	0.1187&	0.3900&	0.0985&	0.1009&	0.1075&	0.3761&	0.0785&	0.0828&	0.0911&	0.3047 \\
		GRU4Rec        & 0.1036&	0.1166&	0.1137&	0.3804&	0.0943&	0.0940&	0.1015&	0.3646&	0.0712&	0.0747&	0.0821&	0.2915 \\
		RUM(I)         & 0.1160&	0.1258&	0.1245&	0.3921&	0.1013&	0.1005&	0.1086&	0.3790&	0.0790&	0.0853&	0.0936&	0.3166 \\
		RUM(F)         & 0.1142&	0.1274&	0.1274&	0.3928&	\underline{0.1041}&	0.1047&	0.1146&	0.3973&	0.0803&	0.0855&	0.0925&	0.3199 \\
		Caser          & 0.1152&	\underline{0.1321}& \underline{0.1456}&	\underline{0.4060}&	0.1038&	\underline{0.1136}&	\underline{0.1252}&	\underline{0.3975}&	0.0819&	0.0873&	0.1029&	0.3378 \\
		SASRec         & \underline{0.1183}&	0.1295&	 0.1436&	0.4054&	0.1026&	0.1105&	0.1241&	0.3922&	\underline{0.0820}&	\underline{0.0875}& \underline{0.1036}&	\underline{0.3384} \\
		\hline
		RNS            & \textbf{0.1329}&	\textbf{0.1531}&	\textbf{0.1648}&	\textbf{0.4446}&	\textbf{0.1146}&	\textbf{0.1252}&	\textbf{0.1388}&	\textbf{0.4362}&	\textbf{0.0894}&	\textbf{0.0943}&	\textbf{0.1120}&	\textbf{0.3614} \\
		Improvement    & 12.4\%&	15.9\%&	13.2\%&	9.5\%&	10.0\%&	10.2\%&	10.9\%&	9.7\%&	9.0\%&	7.8\%&	8.1\%&	6.8\% \\
		\bottomrule
	\end{tabular}}
	\caption{Performance comparison for baselines and \baby. The best and second best results are highlighted in boldface and underlined respectively. Improvement over the best baseline are shown in the last row.}
	\label{tab2}
\end{table*}

We perform our experiments on the Amazon dataset\footnote{http://jmcauley.ucsd.edu/data/amazon/}. This dataset contains product purchase history from Amazon ranging from May 1996 to July 2014. We  conduct experiments on three categories: \textit{Instant Video (IV)}, \textit{Pet Supplies (PS)} and \textit{Tools and Home Improvement (THI)}. Following prevous works~\cite{wsdm18:chen}, the users that have purchased less than $10$ items are removed. We then take explicit ratings as purchase behavior. We hold the first 70\% of items in each user's sequence as the training set and the rest are used for testing. After preprocessing, basic statistics of the datasets are listed in Table~\ref{tab1}.

\paragraph{Metrics.}
Following previous works~\cite{icdm18:kang}, for each test item, we randomly sample 100 negative items, and rank these items with the items in the testing set together. We adopt 4 widely used metrics for performance evaluation: \textit{Precision@N, Recall@N, NDCG@N and HR@N}, where \textit{N} indicates top-$N$ ranked items. The first two metrics regard the recommendation task as classification problem and evaluate the Top-\textit{N} classification results. \textit{NDCG@N} (Normalized Discounted Cumulative Gain) evaluates ranking quality by taking the positions of ground-truth items in the ranking list into consideration. \textit{HR@N} (Hit Ratio) gives the ratio of testing instances that can obtain at least one correct recommendation in the top-$N$. 

\paragraph{Baseline Methods.}
We compare \baby with the following state-of-the-art baselines including sequential learning based models and neural network based models: (1) the popularity based recommender, \textbf{Pop}; (2) the factor  based sequential pattern learning, \textbf{FPMC}~\cite{www10:rendle}; (3) RNN based sequential recommnder, \textbf{GRU4Rec}~\cite{corr15:hidasi}; (4) attention and memory network based sequential recommnder, item-level \textbf{RUM(I)} and feature-level \textbf{RUM(F)}~\cite{wsdm18:chen}; (5) CNN based sequential recommder, \textbf{Caser}~\cite{wsdm18:tang}; (6) self-attention based sequential recommnder, \textbf{SASRec}~\cite{icdm18:kang}.

\paragraph{Parameter Settings.}
As to the baselines, we utilize the recommended setting by their original work. For the proposed \baby, we set $L=5$ which is a common setting in most relevant works. The number of aspects is set to be $5$\footnote{We observe that $K$ is optimal at $5$ on all datasets. Detailed results are not included due to page limitation.}, and embedding size is set to $25$, $n=10$ and $h=\{1,3,5,7,9\}$.  The learning rate is set to $0.001$, $x=3,\alpha=0.1$ and $\lambda=0.0001$. We set $N=5$ for performance evaluation.

%============================================
\subsection{Results and Discussion}
%============================================
The overall performance of different methods is presented in Table~\ref{tab2}. Several observations can be made:

First, the non-personalized Pop method gives the worst performance on all metrics across three datasets, which is followed by RNN based solution GRU4Rec. Also, The non-neural method FPMC surpasses GRU4Rec significantly. Here FPMC models user preference through integrating matrix factorization (MF) with sequential information, while GRU4Rec utilizes RNNs based on the hypothesis that the latter state depends on the previous one. This result might highlight that using RNN along is not sufficient to learn users' dynamic preference and long-term user preference modeled by MF in FPMC is also beneficial.
	
Second, RUM(I) and RUM(F) achieve comparable results on three datasets, indicating the memory mechanism and updating operation in RUM can provide expressive ability to model user dynamic preference. Caser and SASRec outperform the other baselines on all settings except \textit{Precision@5} in \textit{Pet Supplies}. Caser generally achieves better recommendation performance than other baselines in both \textit{Instant Video} and \textit{Pet Supplies}. In contrast, SASRec domaintes in \textit{Tools and Home Improvement}.
	 
Third, our proposed \baby consistently outperforms all baselines by a wide margin across all datasets. Specifically, the average performance gains of all metrics on three datasets are $12.8\%$, $10.2\%$, $7.9\%$ in \textit{Instant Video}, \textit{Pet Supplies} and \textit{Tools and Home Improvement} respectively. The substantial improvement of our model over the baselines could be credited to the following reasons: (1) we model user's long-term preference through aspect-aware convolutional processing of the review document. The rich semantics provided by the reviews help us better characterize user's general preference; (2) we capture sequential patterns at both union-level and individual-level based on a novel hierarchical attention over attention mechanism. These traits are confirmed by further analysis of \baby in the following.    

%============================================
\subsection{Ablation Study}
%============================================
\begin{table}
	\scriptsize
	\begin{tabular}{l|cc|cc|cc}
			\toprule
			Datasets & \multicolumn{2}{c}{IV} & \multicolumn{2}{c}{PS} & \multicolumn{2}{c}{THI} \\
			\midrule
			Measures@5 & Recall & HR & Recall & HR & Recall & HR \\
			\hline
			RNS    	   & \textbf{0.1531}&	\textbf{0.4446}&	\textbf{0.1252}&	\textbf{0.4362}&	\textbf{0.0943}&	\textbf{0.3614} \\
			RNS-i      & 0.1455&	0.4264&	0.1214&	0.4277&	0.0904&	0.3487 \\
			RNS-u      & 0.1480&	0.4308&	0.1219&	0.4297&	0.0885&	0.3446 \\
			RNS-pe     & 0.1516&	0.4359&	0.1220&	0.4289&	0.0905&	0.3489 \\
			RNS-at     & 0.1468&	0.4176&	0.1214&	0.4293&	0.0926&	0.3544 \\
			\bottomrule
	\end{tabular}
	\caption{Ablation study of RNS.}
	\label{tab3}
\end{table}

As there are many components in \baby architecture, we evaluate their effects via ablation study, as shown in Table~\ref{tab3}. Due to space limitation, we only list the results with respect to \textit{Recall} and \textit{HR}. Similar performance patterns are also observed for \textit{Precision} and \textit{NDCG}. RNS-u means that
the short-term user preference is modeled by excluding union-level, \ie $\textbf{p}^{s}_u = \textbf{p}^{s_2}_u$, and RNS-i for excluding  individual level, \ie $\textbf{p}^{s}_u = \textbf{p}^{s_1}_u$. RNS-pe means that position embeddings are excluded, while RNS-at means that only one aspect is considered, \ie without applying aspect transformation.

From Table~\ref{tab3}, we can observe that keeping only union-level or individual-level preference will reduce the recommendation accuracy, suggesting that the hierarchical attention mechanism is beneficial to identify the user's dynamic preference. On both \textit{Instant Video} and \textit{Pet Supplies} datasets, RNS-u outperforms RNS-i on two metrics, while on \textit{Tools and Home Improvement} dataset RNS-i is better. This observation suggests that the importance of union-level and individual-level depends on dataset. Without the position embeddings, RNS-pe achieves worse performance than \baby on three datasets. This suggests that temporal information in forms of purchase order is a useful signal for sequential recommendation. Also, without utilizing aspect-aware feature extraction, \baby experiences the performance degradation to some extent. This suggests that aspect-aware features extracted from reviews are more discriminative.

%============================================
\subsection{Parameter Sensitivity}
%============================================
\begin{figure}
	\centering
	\subfigure{
		\includegraphics[height=1.25in, width=1.68in]{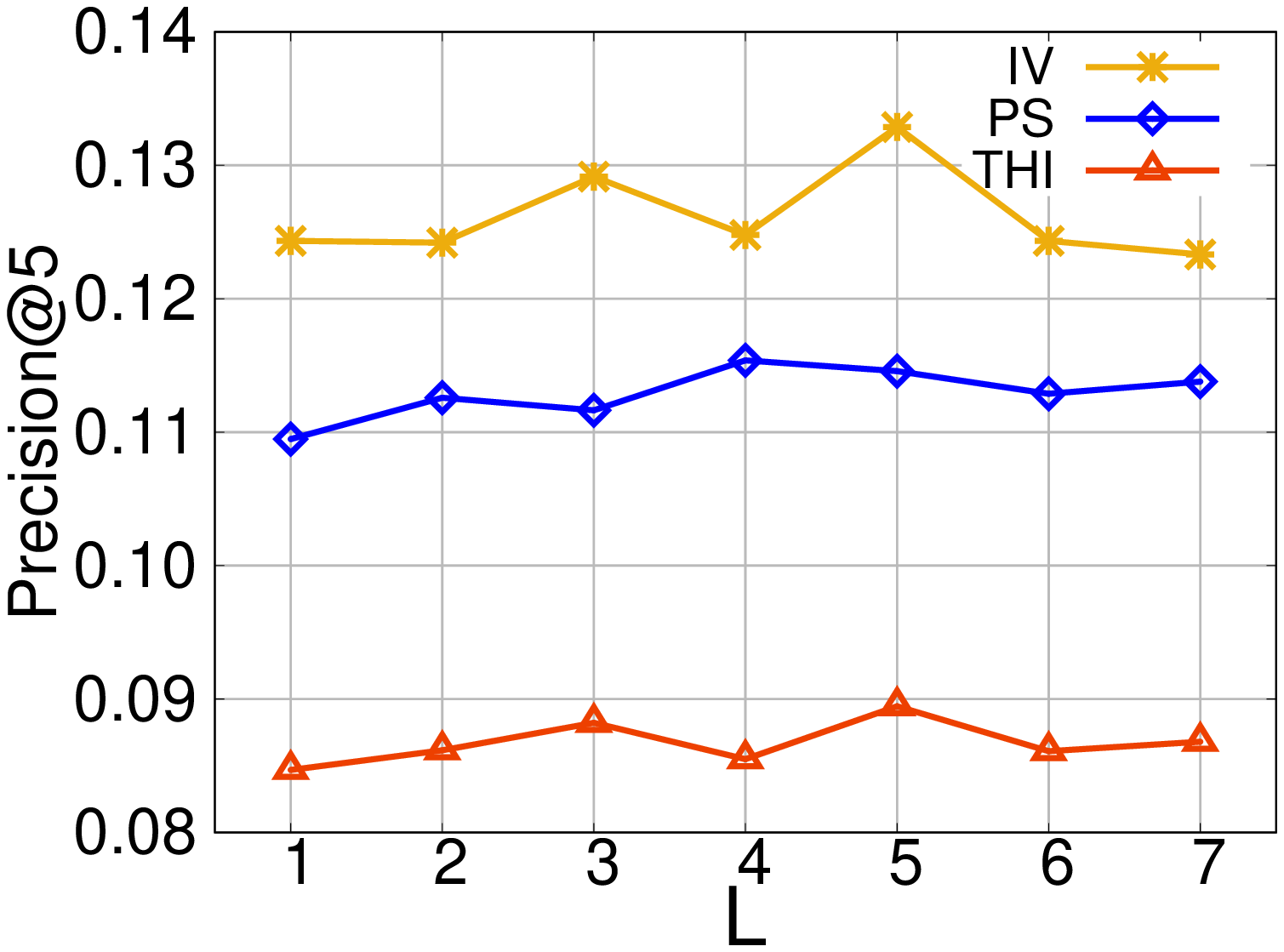}
		\label{figureL}
	}
	\hspace{-0.2in}
	\subfigure{
		\includegraphics[height=1.25in, width=1.68in]{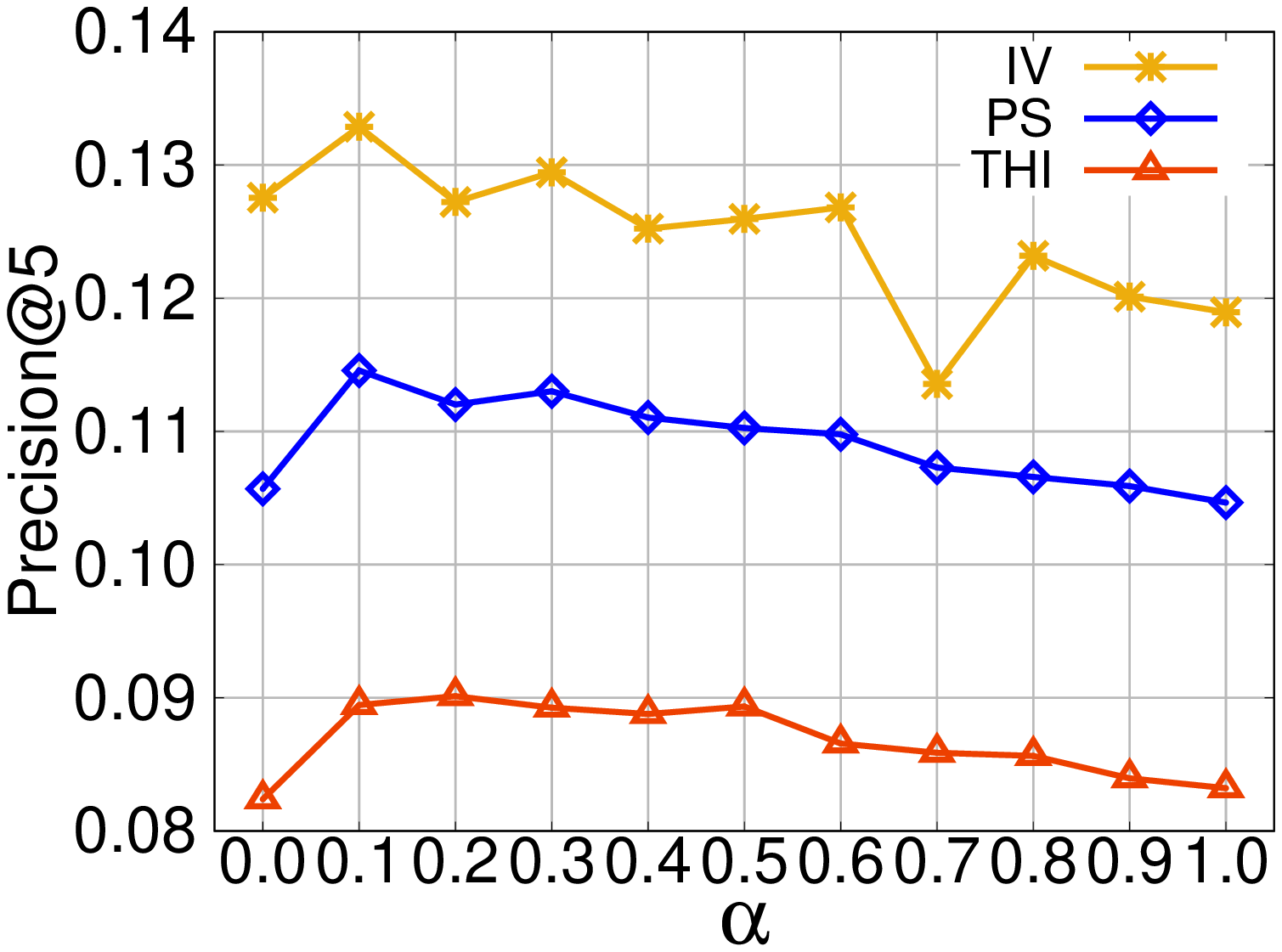}
		\label{figurealpha}	
	}
	\caption{Parameter sensitivity of RNS: $L$ and $\alpha$}
	\label{figure2}
\end{figure}

We further study the effects of parameters $L$ and $\alpha$ on model performance. The results of \textit{Precision@5} on three datasets are shown. We omit the similar patterns observed for other metrics due to space limitation. Figure~\ref{figureL} plots the performance pattern by varying $L$ values. We observe that \baby consistently achieves better performance at $L=5$ across datasets. When $L$ becomes smaller or larger, \baby experiences performance degradation to some extent. This is reasonable since a small $L$ produces less useful historical signals and a large $L$ would inevitably introduce much noise. Hence, we set $L=5$ in our experiments.

Parameter $\alpha$ in Equation~\ref{equ1} controls the importance of sequential patterns (\ie union-level and individual-level preferences) in \baby. Figure~\ref{figurealpha} plots the performance pattern by varying $\alpha$ values. When $\alpha=0$, \baby degrades to a variant of DeepCoNN proposed for rating prediction~\cite{wsdm17:deepconn}. We observe that \baby performs much worse in this setting. This is consistent with the existing finding that sequential patterns are crucial for sequential recommendation~\cite{www10:rendle}. The optimal performance is consistently achieved when $\alpha=0.1$ in all the three datasets. When $\alpha$ becomes increasingly larger, the performance of \baby further deteriorates to a larger extent. These results further suggest that both short-term and long-term user preferences are complementary to each other for sequential recommendation. Accordingly, we set $\alpha=0.1$ in the experiments.

\subsection{Case Study}
\begin{figure}
	\centering
	\subfigure[]{
		\includegraphics[height=1.32in, width=1.58in]{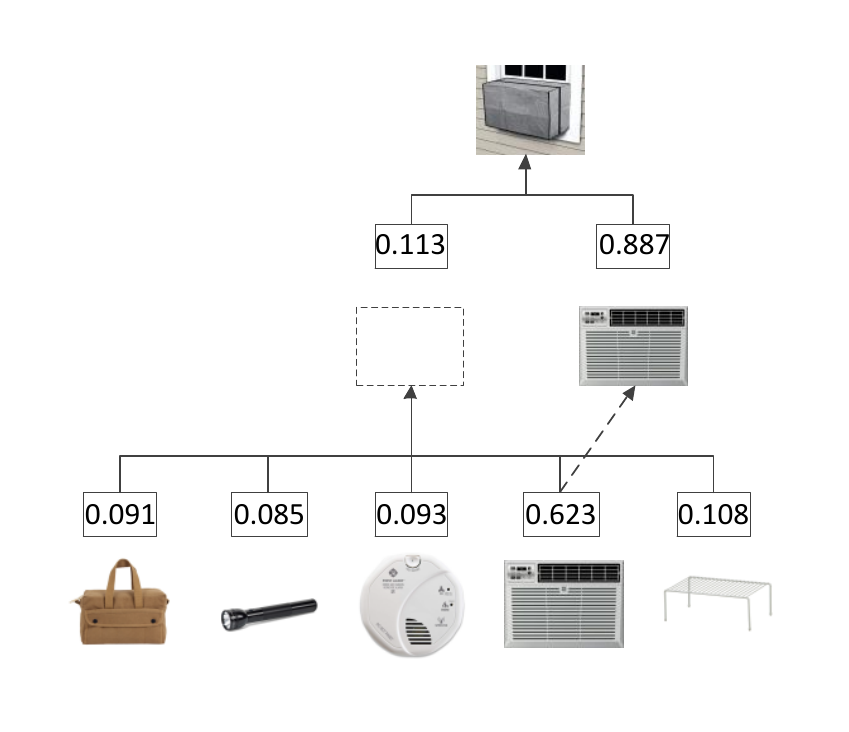}
		\label{case1}
	}
	\subfigure[]{
		\includegraphics[height=1.32in, width=1.58in]{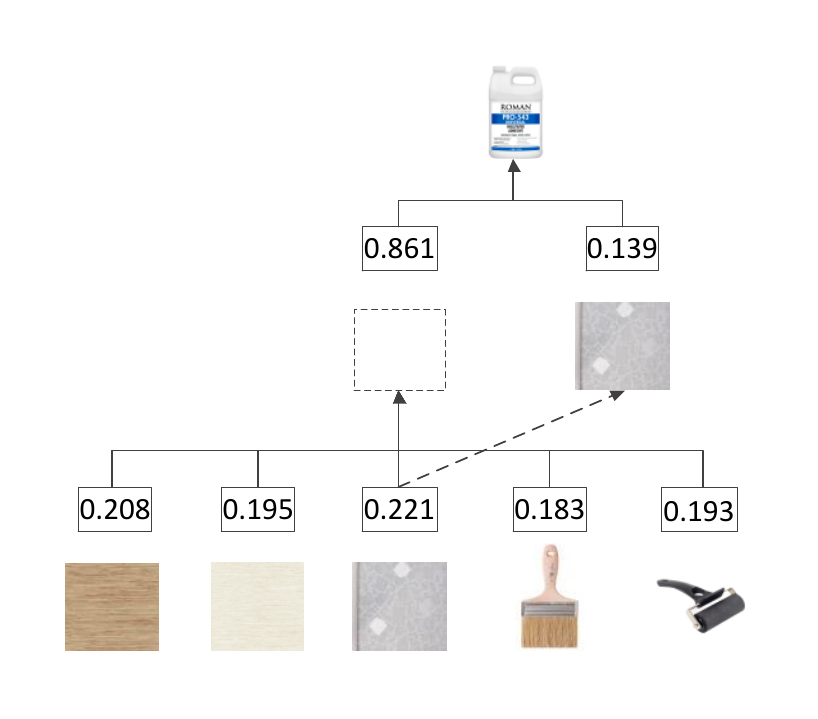}
		\label{case2}	
	}
	\caption{Illustration of attention weights hierarchically.}
	\label{figure3}
\end{figure}

To illustrate the intuition of the hierarchical attention over attention at both union-level and individual-level, We present the case study about two users sampled in \textit{Tools and Home Improvement}. The attention weights in both union-level and individual-level are displayed for interpretation. 

Figure~\ref{case1} demonstrates a case where individual level is more crucial than union-level, as the user wants to buy an air conditioner cover just because he bought an air conditioner not long ago. Figure~\ref{case2} demonstrates a completely opposite example, where all recently purchased 5 items contribute nearly equal for next purchase decision. The user needs wallpaper adhesive since he bought several kinds of wallpapers and related tools before. These observations confirm that the proposed hierarchical attention over attention mechanism in \baby is effective to capture users' dynamic preference.
%============================================
\section{Conclusion}\label{sec:conclusion}
%============================================
In this paper, we propose a novel neural model for sequential recommendation with reviews, named \baby. Our model incorporates both a user's long-term intrinsic preference and short-term preference to predict her next actions. It utilizes aspect-aware CNN network to extract user and item representations from the corresponding review documents. A novel hierarchical attention over attention mechanism is proposed to capture sequential patterns at both union-level and individual-level. Experimental results show that \baby significantly outperforms strong state-of-the-art methods under different metrics. We will extend \baby to facilitate fine-grained textual explanation for sequential recommendation.

\section*{Acknowledgements} This research was supported by National Natural Science Foundation of China (No.~61872278, No.~U1636219, No.~U1804263), National Key R\&D Program of China (No.~2016YFB0801303, No.~2016QY01W0105), Science and Technology Innovation Talent Project of Henan Province (No. 184200510018).

\bibliographystyle{named}
\bibliography{ijcai19}

\end{document}